\documentclass[twocolumn,tightenlines,aps,preprintnumbers,superscriptaddress,showpacs,floatfix]{revtex4}

\usepackage{graphicx,psfig}
\usepackage{epsfig}
\usepackage{amssymb}
\usepackage{amsmath,amssymb,textcomp,array}

\newcommand{\mn}{{Mon.\@ Not.\@ Roy.\@ Ast.\@ Soc.\ }}
\newcommand{\asta}{{Astron.\@ Astrophys.\ }}
\newcommand{\aj}{{Astron.\@ J.\ }}

\newcommand{ \jetpl}{JETP Lett.\ }
\newcommand{\jcap}{J.~Cosmol.~Astropart.~Phys.~}

\newcommand{\etal}{{et al.}}
\newcommand{\ie}{{i.e.}}

\newcommand{\eg}{{e.g.,~}}

\newcommand{\beq}{\begin{equation}}
\newcommand{\eeq}{\end{equation}}
\newcommand{\ber}{\begin{eqnarray}}
\newcommand{\eer}{\end{eqnarray}}
\newcommand{\lleq}{\lower0.9ex\hbox{ $\buildrel < \over \sim$} ~}
\newcommand{\ggeq}{\lower0.9ex\hbox{ $\buildrel > \over \sim$} ~}
\newcommand{\lsim}{\ \lower-1.5pt\vbox{\hbox{\rlap{$<$}\lower5.3pt\vbox{\hbox{$\sim$}}}}\ }
\newcommand{\gsim}{\ \lower-1.5pt\vbox{\hbox{\rlap{$>$}\lower5.3pt\vbox{\hbox{$\sim$}}}}\ }

\newcommand{\de}{dark energy}
\newcommand{\cc}{{cosmological constant}}
\newcommand{\adot}{\dot{a}}
\newcommand{\addot}{\ddot{a}}
\newcommand{\atridot}{\stackrel{...}{a}}
\newcommand{\phidot}{\dot{\phi}}
\newcommand{\phiddot}{\ddot{\phi}}
\newcommand{\omt}{\Omega_{0 \rm m}}

\newcommand{\omde}{\Omega_{DE}}

\newcommand{\omm}{{\rm Om}}

\newcommand{\w}{w_{DE}}
\newcommand{\half}{{1\over 2}}

\begin{document}

\title{Reconstructing Dark Energy : A Comparison of Cosmological Parameters}

\author{Alexander V. Pan}
\affiliation{Purdue University, West Lafayette, IN 47907, USA}
\affiliation{ISR-1, ISR Division, Los Alamos National Laboratory, Los Alamos, NM 87545, USA}
\author{Ujjaini Alam}
\affiliation{ISR-1, ISR Division, Los Alamos National Laboratory, Los Alamos, NM 87545, USA}

\thispagestyle{empty}

\sloppy

\begin{abstract}
\small{ 
A large number of cosmological parameters have been suggested for
obtaining information on the nature of dark energy. In this work, we study the
efficacy of these different parameters in discriminating theoretical
models of dark energy, using both currently available supernova (SNe) data,
and simulations of future observations. We find that the current data
does not put strong constraints on the nature of dark energy, irrespective of
the cosmological parameter used. For future data, we find that the
although deceleration parameter can accurately reconstruct some
dark energy models, it is unable to discriminate between different models of
dark energy, therefore limiting its usefulness. Physical parameters such as
the equation of state of dark energy, or the dark energy density do a good job of both
reconstruction and discrimination {\it if} the matter density is known
to high accuracy. However, uncertainty in matter density reduces the
efficacy of these parameters. A recently proposed parameter, ${\rm Om}$,
constructed from the first derivative of the SNe data, works very well
in discriminating different theoretical models of dark energy, and has the
added advantage of not being dependent on the value of matter
density. Thus we find that a cosmological parameter constructed from
the first derivative of the data, for which the theoretical models of
dark energy are sufficiently distant from each other, and which is independent
of the matter density, performs the best in reconstructing dark energy from
SNe data. }
\end{abstract}

\maketitle

\flushbottom

\section{Introduction}\label{intro}

The nature of dark energy is one of the most tantalizing mysteries in
cosmology today.  Observations of high redshift type Ia supernovae
tell us that the expansion of the universe is accelerating at present
\cite{union2}, which can not be satisfactorily explained in the standard
cold dark matter (CDM) scenario. Independent observations of the
cosmic microwave background \cite{wmap7} and large scale structure
\cite{sdss} tell us that two-thirds of the present density of the
universe is composed of some unknown component. The study of this
unknown ``\de'' is of great interest among cosmologists today.

Various theoretical models for \de~have been suggested, the simplest
being the \cc~model with constant dark energy density and equation of
state $\w = -1$. Other models of dark energy include physically
motivated models like the scalar field quintessence and Chaplygin gas
models, as well as geometrically motivated models like scalar-tensor
theories and higher dimensional braneworld models (see \cite{de} and
references therein).

The growing number of theoretical \de~models has inspired a
complementary, data-driven approach in which the properties of \de~are
reconstructed from the data by studying the cosmological parameters
characterizing \de. Two primary methods are used for reconstructing
cosmological parameters. In the first approach, known as parametric
reconstruction, a sufficiently general fitting function is used to
represent the parameter in the analysis. This suffers from the
possibility of bias, depending on the form chosen for the
parameter. The second method is that of non-parametric reconstruction,
in which no specific form is assumed for the parameter. The difficulty
with this is that the parameters of interest are usually obtained by
taking the first or second derivative of the data, therefore, direct
reconstruction involving differentiation of noisy data can lead to
large errors.  Many different cosmological parameters have been
suggested both these reconstruction methods (see \cite{parm} and
references therein). In this work, we attempt to study the relative
efficacy of the different cosmological parameters in reconstructing
\de~from observations, and discriminating between different
theoretical \de~models, using the parametric reconstruction formalism.

The paper is arranged as follows-- section~\ref{meth} contains a
description of the data and methods used in the analysis,
section~\ref{res} outlines the results, and section~\ref{concl}
presents the conclusions.

\section{Methodology}\label{meth}

This work attempts to classify the different cosmological parameters
that characterize \de~in terms of their efficiency in constraining the
nature of \de. In order to do this, we analyze cosmological
observations using the different parameters to compare how accurately
these parameters reconstruct and discriminate between various
theoretical models of dark energy. In this analysis, we primarily use
Type Ia supernova data along with information on the present day
matter content of the universe.

\subsection{Supernova Data}

Type Ia supernova are the most direct evidence for the existence of
dark energy at present. From early twentieth century, they were
investigated as standard candles and many attempts were made to use
them to measure the Hubble parameter and the deceleration of the
universe \cite{early_sne}. The first cosmologically significant
results for deceleration of the universe were produced in the late
nineties, when two observational groups \cite{riess, perl}
independently showed that the expansion of the universe was
accelerating. Since then, there have been numerous other SNe surveys
\cite{sne, union2}, and despite being plagued by systematics, these
remain our best observational tool for studying dark energy.

Supernova data by itself is insufficient to break the degeneracy
between dark energy parameters and the curvature of the
universe. Since the objective of this paper is to constrain dark
energy using SNe data, we restrict our analysis to a flat model
($\Omega_{\kappa} = 1 $) of the universe which is favoured by the
current CMB data \cite{wmap7}. The data is in the form
\beq 
\mu_B(z) = 5 {\rm log}_{10} d_L(z) + {\cal M} \,\,, 
\eeq 
where ${\cal M}$ represents a noise parameter usually marginalized
over, and the luminosity distance $d_L(z)$ is related to the
cosmological parameters in a flat universe as--
\ber 
d_L(z) &=& c(1+z) \int_0^z \frac{dz}{H(z)} \\ 
H(z) &=& \frac{\adot}{a} = H_0 \sqrt{\omt (1+z)^3 + \omde(z)} \,\,.  
\eer 
($a$ is the scale factor representing the expansion of the universe,
$H(z)$ is the Hubble parameter, $c$ denotes the speed of light and
$H_0$ the present value of the Hubble parameter in km/s/Mpc).

We perform a maximum likelihood analysis on the supernova data to
obtain constraints on the various \de~parameters, the likelihood being
defined as
\ber
{\cal L} &\propto& e^{-\chi^2/2} \\
\chi^2 &=& \sum_{i=1}^{\rm N_{data}} \left( \frac{\mu_{B,i}(z_i)-\mu_B(z_i;{\cal M}, p_j)}{\sigma_{\mu_B, i}} \right)^2 \,\,,
\eer
where $p_j$ are the parameters of the fitting function chosen to
represent the cosmological parameter being studied.

We use one of the most current SNe datasets \cite{union2} for our
analysis. However, as we shall see, the current data is not yet of
such a quality that it could strongly discriminate between different
models of \de. We therefore also simulate three datasets based on our
expectations from future surveys \cite{jdem}, to study the information
that could be obtained from future data. The datasets used in our
analysis are--

\begin{itemize}

\item
Dataset I : 
Currently available Union2 dataset, with $\sim 550$ SNe between
redshifts $z = 0-1.4$, and average statistical error of
$\sigma_{\mu_B} \sim 0.1-0.3$ mags.

\item
Dataset II A :
Simulated dataset based on future JDEM-like SNe surveys containing
$\sim 2000$ SNe distributed over a redshift range of $z = 0-1.7$ with
a larger concentration of supernovae in the midrange redshift bins ($z
= 0.4-1.1$) and average statistical errors of $\sigma_{\mu_B}= 0.13$
mags \cite{jdem}.  The theoretical model used to simulate the data is
the \cc~model with $H_0 = 72~{\rm km/s/Mpc}$, the matter density $\omt
= 0.27$, and the equation of state of \de~$\w = -1$.

\item 
Dataset II B :
Simulated dataset based on a JDEM-like survey as in II A, using a
theoretical model of quintessence with a minimally coupled scalar
field whose equation of motion is given by 
\beq\label{scal} 
\phiddot + 3 H \phidot+ \frac{dV}{d\phi} = 0 \,\,, 
\eeq 
with a potential \cite{rp} 
\beq\label{rp} 
V(\phi) = V_0 \phi^{\alpha}~; \ \alpha = 2 \,\,, 
\eeq 
and the equation of state of \de 
\beq\label{model2} 
\w = \frac{\half\phidot^2 - V(\phi)}{\half\phidot^2 + V(\phi)} \,\,.  
\eeq
$\omt$ and $H_0$ are the same as in Dataset II A.

\item 
Dataset II C :
Simulated JDEM-like dataset using a variable \de~model with the
equation of state given by \cite{coras}
\ber
\w(z) =&& w_0+(w_m-w_0))\frac{1+e^{\frac{1}{\Delta_t (1+z_t)}}}{1-e^{\frac{1}{\Delta_t}}}  \nonumber\\
&& \times \left[ 1- \frac{e^{\frac{1}{\Delta_t}} + e^{\frac{1}{\Delta_t (1+z_t)}}}{e^{\frac{1}{\Delta_t (1+z)}} +e^{\frac{1}{\Delta_t (1+z_t)}}} \right] \,\,,
\eer
with the values $w_0 = -1.0,~w_m = -0.5,~z_t = 0.5,~\Delta_t =
0.05$. This model has $\w \geq -1$ everywhere, and may be realized by a
standard quintessence field. $\omt$ and $H_0$ are the same as in Dataset
II A.

\end{itemize}

\subsection{Parameters of interest}

Several parameters have been suggested in the literature for
reconstructing the nature of \de. These can be broadly divided into
two categories--

\begin{itemize}

\item
Geometrical parameters of \de : 
These are parameters that can be constructed directly from the scale
factor $a$ and its time derivatives. Examples are the Hubble parameter
$H(z)$, and the quantity $\omm (z)$ \cite{omparm}, both constructed
from the first derivative of the supernova data--
\ber
H(z) &=& \frac{\adot}{a}~~,  \\
\omm (z) &=& \frac{H^2(z)/H^2_0 -1}{(1+z)^3 - 1} \,\,,
\eer
the deceleration parameter $q(z)$, constructed from the second
derivative of the data--
\beq
q(z) = -\frac{\addot a}{\adot^2} = -1 + \frac{d{\rm log}H}{d{\rm log}(1+z)} \,\,,
\eeq
and the Statefinder parameters \cite{state}, constructed from the
third derivative of the data--\
\ber
r(z) &=& \frac{\atridot a^2}{\adot^3} \\
s(z) &=& \frac{r - 1}{3 (q - \frac{1}{2})} \,\,.
\eer

\item
Physical parameters of \de :
These parameters, in addition to the scale factor and its derivatives,
also contain physical information such as the matter density. Examples
are the \de~density normalized to the critical energy density
$\rho_{0c} = 3 H_0^2/8 \pi G$, which we denote as $\omde(z)$,
constructed from the first derivative of the data--
\beq
\omde(z) = \frac{H^2}{H_0^2} - \omt (1+z)^3 \,\,,
\eeq
and the equation of state of \de~$\w(z)$, constructed from the second
derivative of the data--
\ber
\w(z) &=& \left. \frac{p}{\rho} \right\vert_{DE} \nonumber\\
&=& \frac{2 (1+z)/3 \ (d{\rm ln}~H/dz) - 1}{1-(H_0/H)^2 \omt (1+z)^3} \,\,.
\eer
The disadvantage for these parameters is their dependence on the
physical nature of \de, which means that for non-physical models of
\de~which do not follow the standard Einstein equations (such as
modified gravity models, or DM+DE interacting models), these
parameters may not be very well-defined. Also, these parameters depend
on the matter density, and therefore their reconstruction may have
added bias due to uncertainty in our knowledge of the matter density.

\end{itemize}

\begin{figure*}
\begin{center}
$\begin{array}{c@{\hspace{1in}}c}
\epsfxsize=2.4in
\epsffile{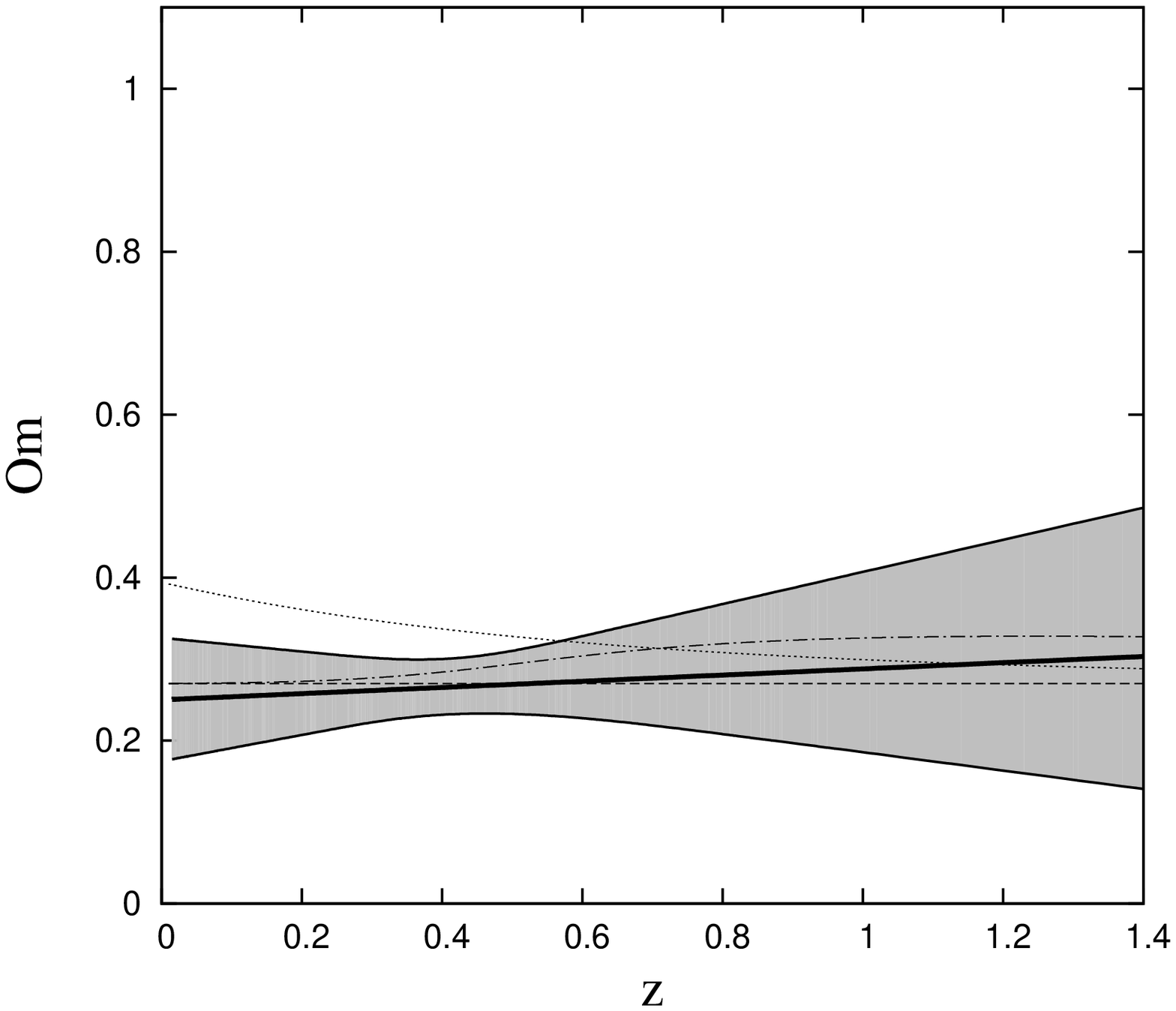} &
	\epsfxsize=2.4in
	\epsffile{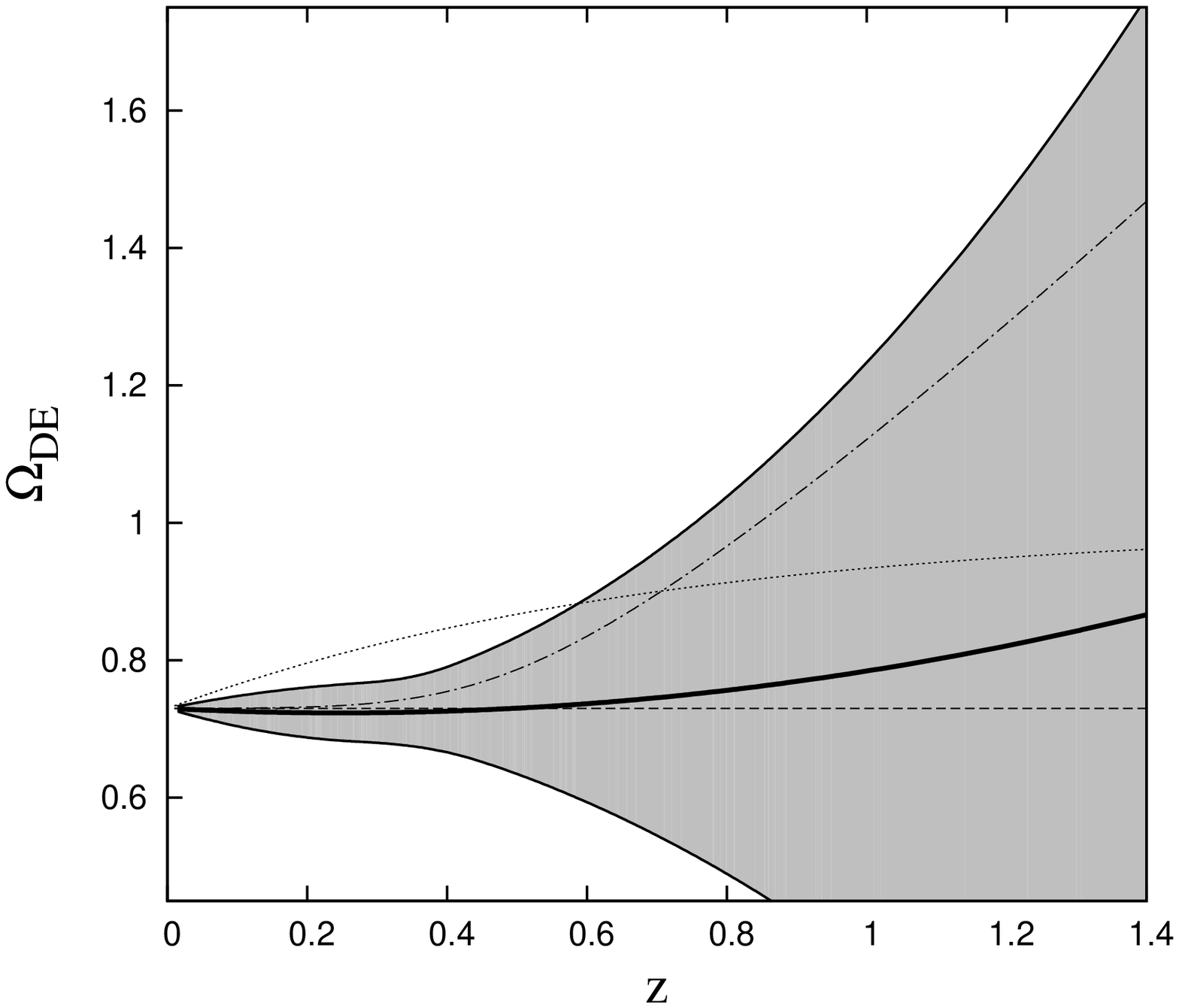} \\ [0.4cm]
\mbox{\bf (a)} & \mbox{\bf (b)}
\end{array}$
$\begin{array}{c@{\hspace{1in}}c}
\epsfxsize=2.4in
\epsffile{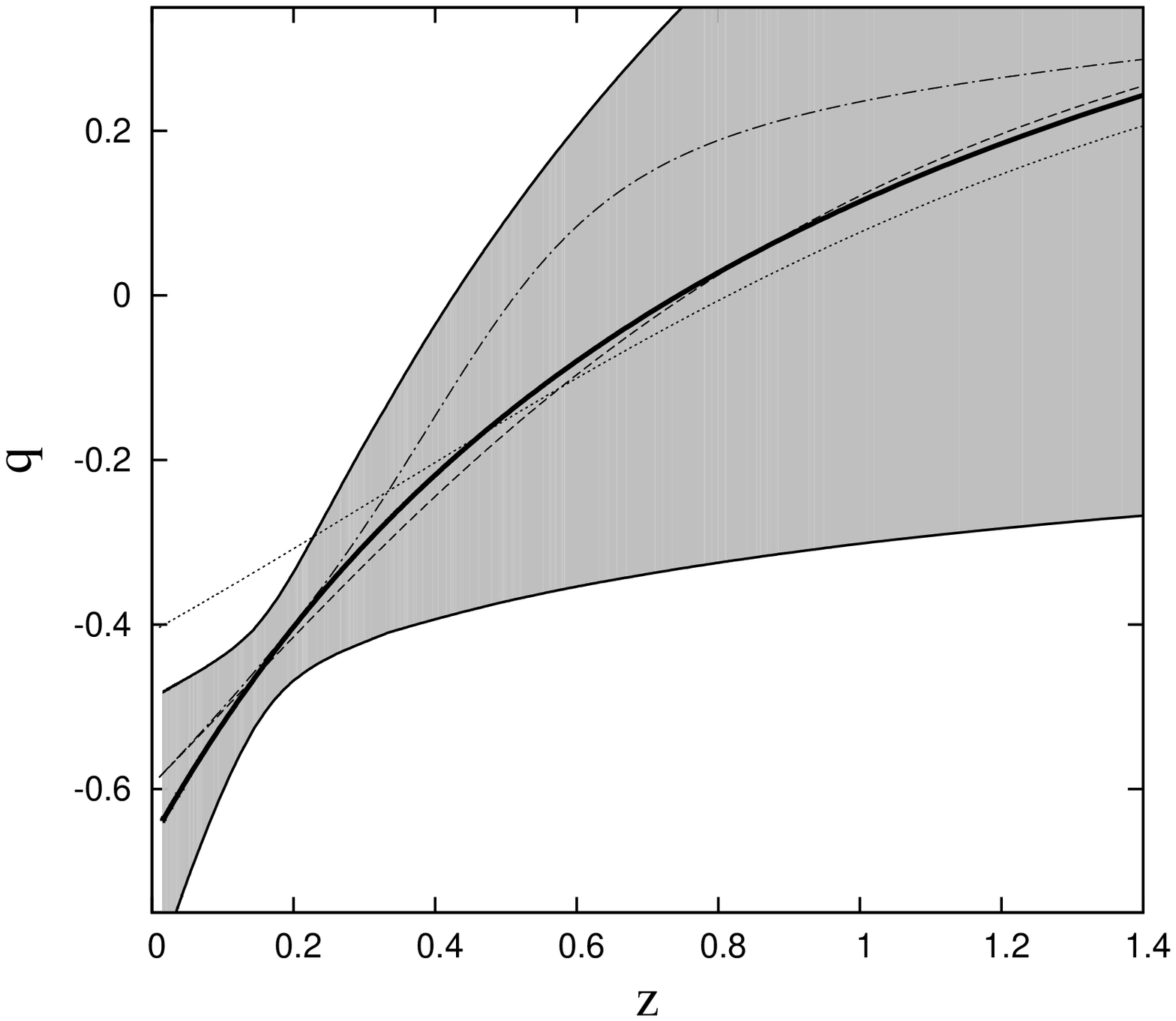} &
	\epsfxsize=2.4in
	\epsffile{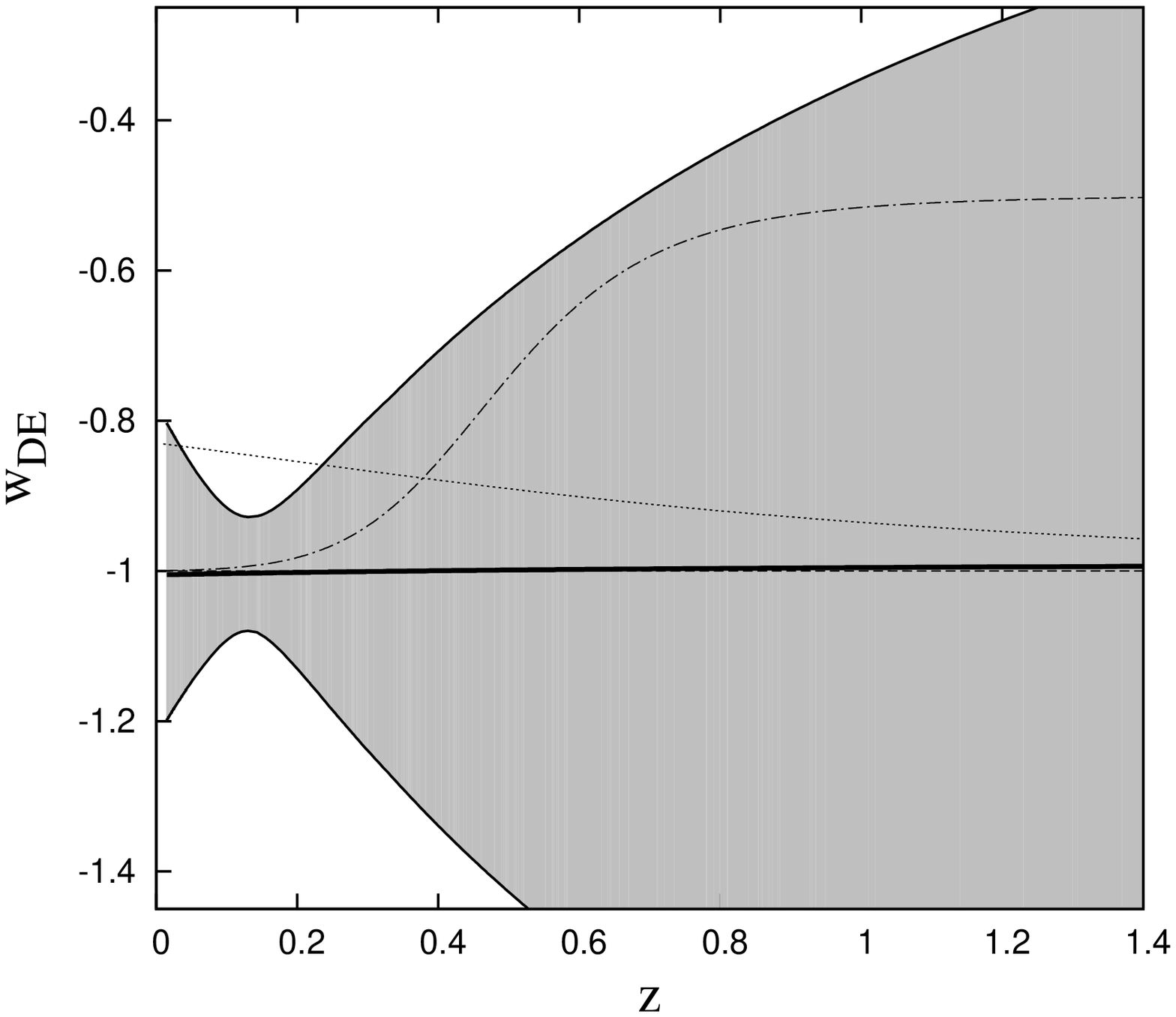} \\ [0.4cm]
\mbox{\bf (c)} & \mbox{\bf (d)}
\end{array}$
\end{center}
\caption{ 
Variation of different cosmological parameters with redshift for
dataset I (real data). Panel (a) shows the quantity $\omm(z)$, panel
(b) shows $\omde(z)$, panel (c) shows $q(z)$ and panel (d) shows
$\w(z)$. The thick solid lines represent the best-fit, and the solid
grey contours represent the $2\sigma$ confidence levels. The dashed,
dotted and dot-dashed lines represent the true model for datasets II
A, B and C respectively. In panels (b) and (d), the solid grey
contours represent the $2\sigma$ confidence levels when $\omt$ is
known exactly.  }
\label{fig:dat1}
\end{figure*}

\begin{figure*}
\begin{center}
$\begin{array}{c@{\hspace{1in}}c}
\epsfxsize=2.4in
\epsffile{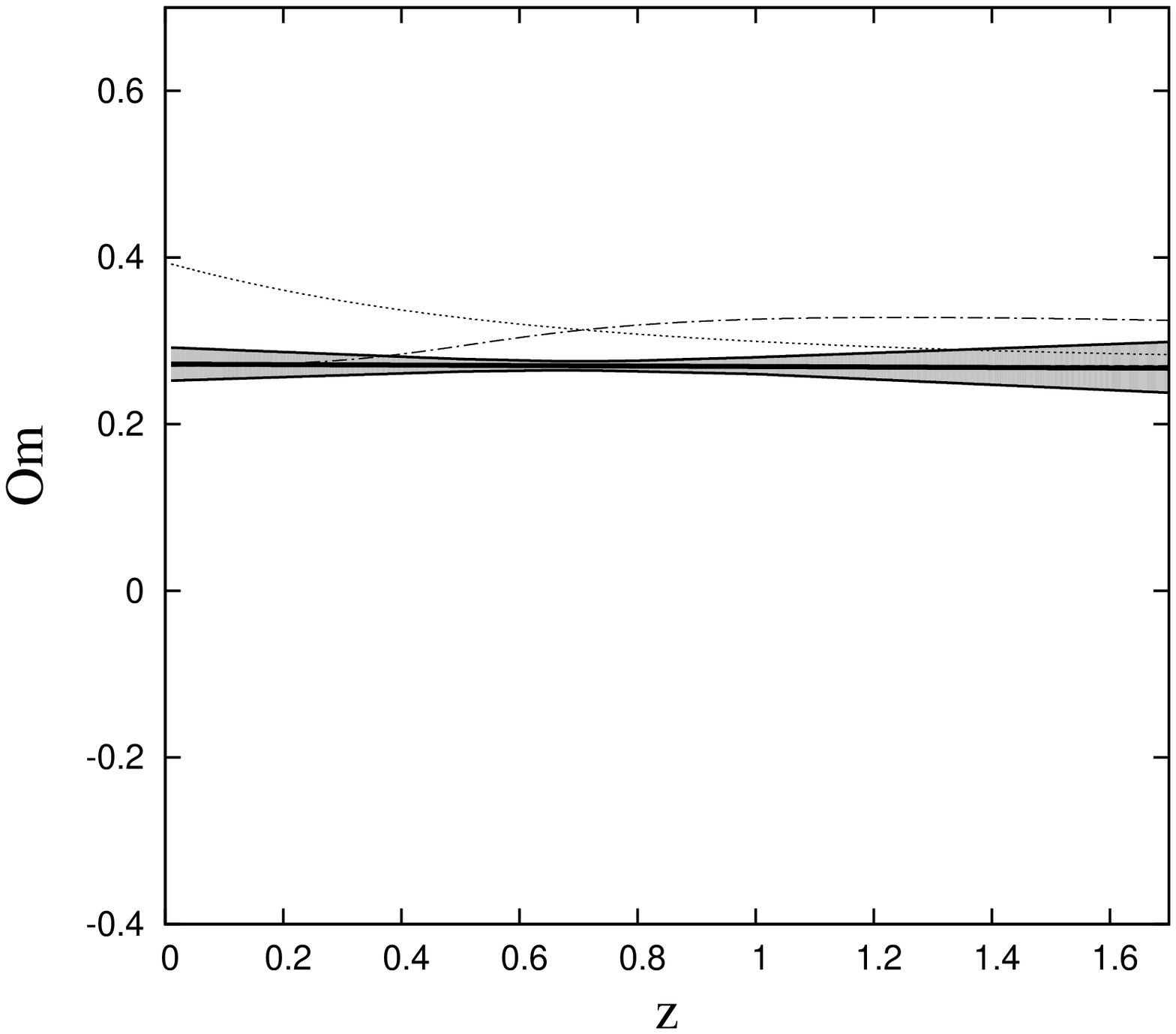} &
	\epsfxsize=2.4in
	\epsffile{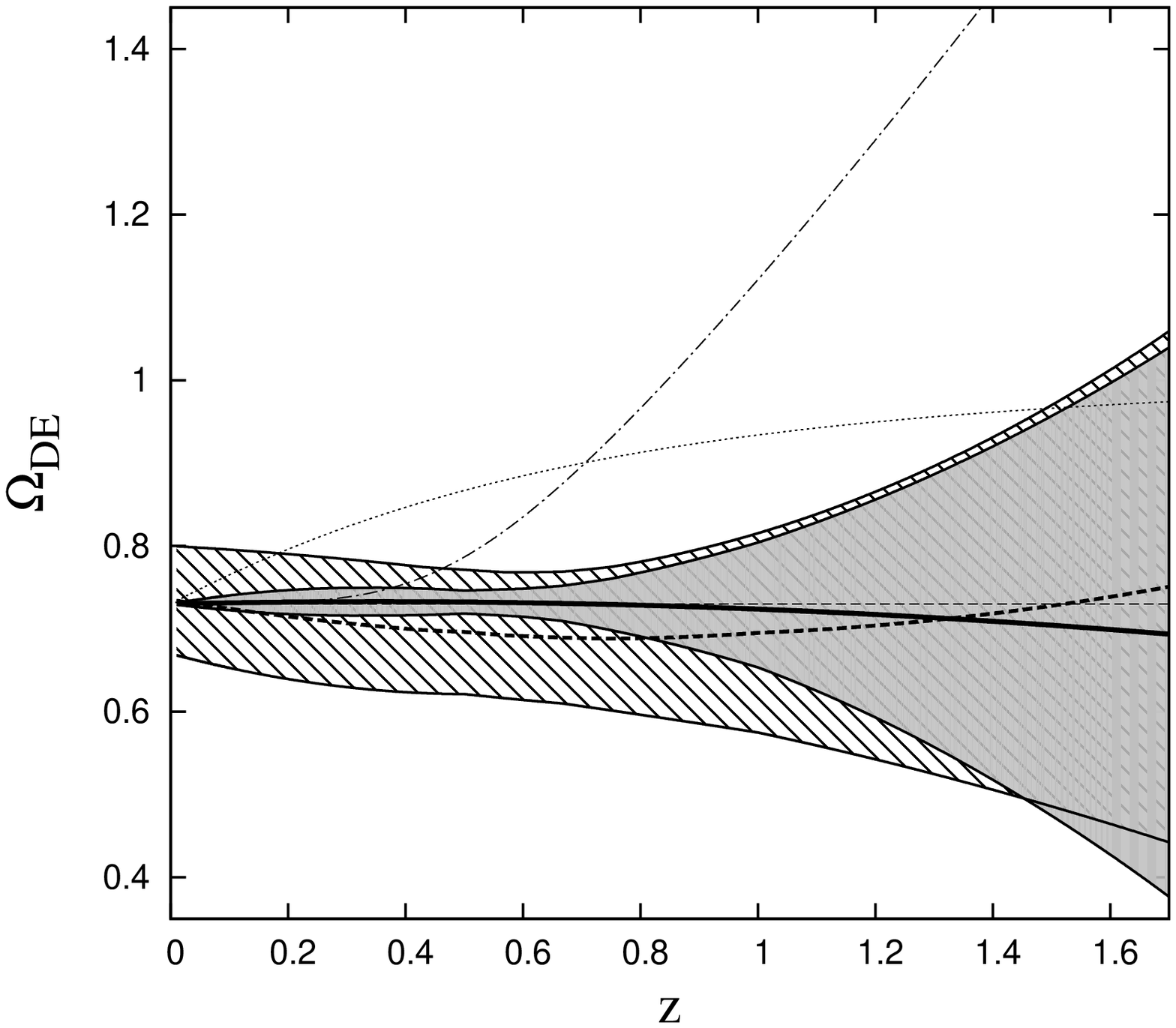} \\ [0.4cm]
\mbox{\bf (a)} & \mbox{\bf (b)}
\end{array}$
$\begin{array}{c@{\hspace{1in}}c}
\epsfxsize=2.4in
\epsffile{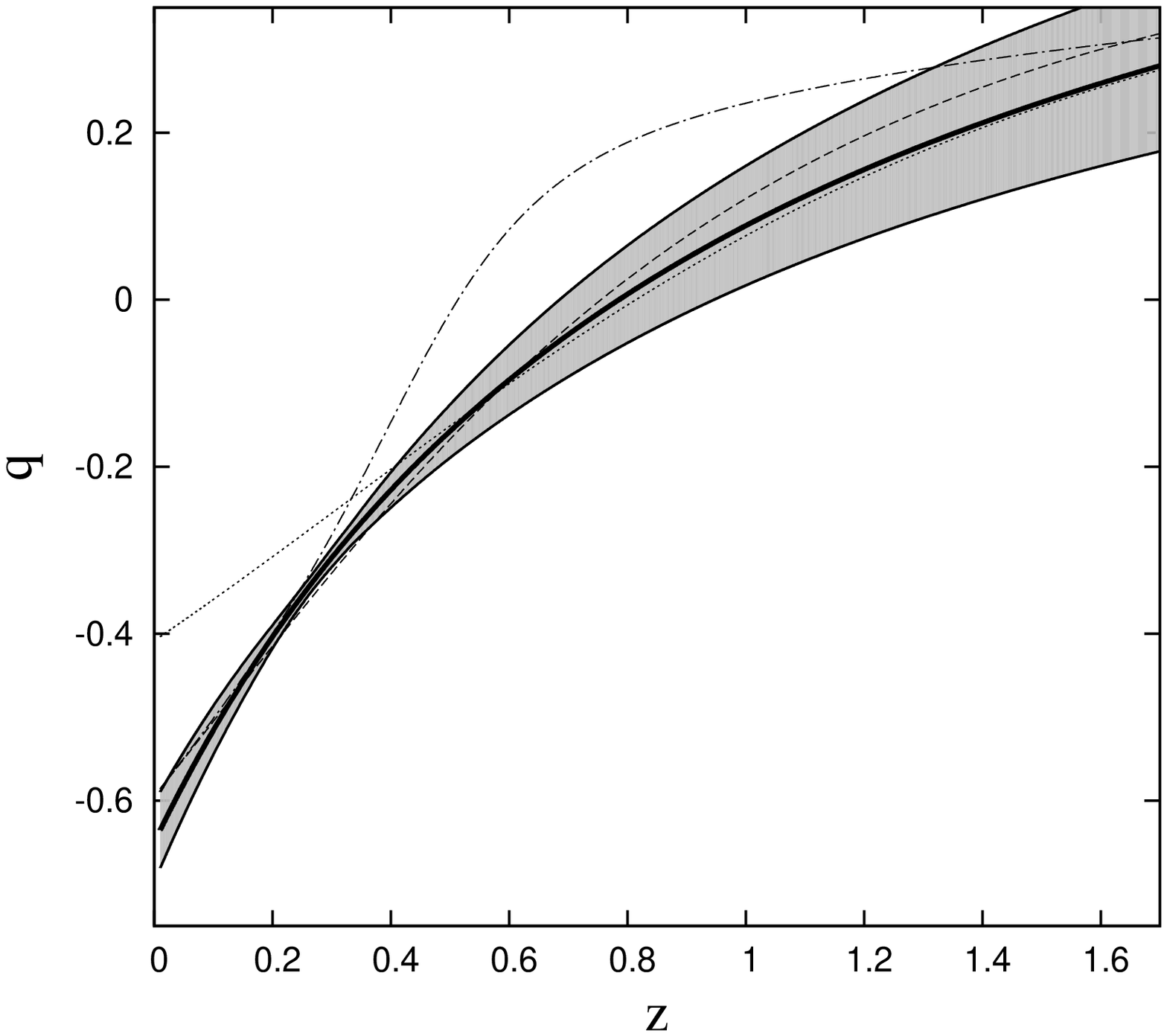} &
	\epsfxsize=2.4in
	\epsffile{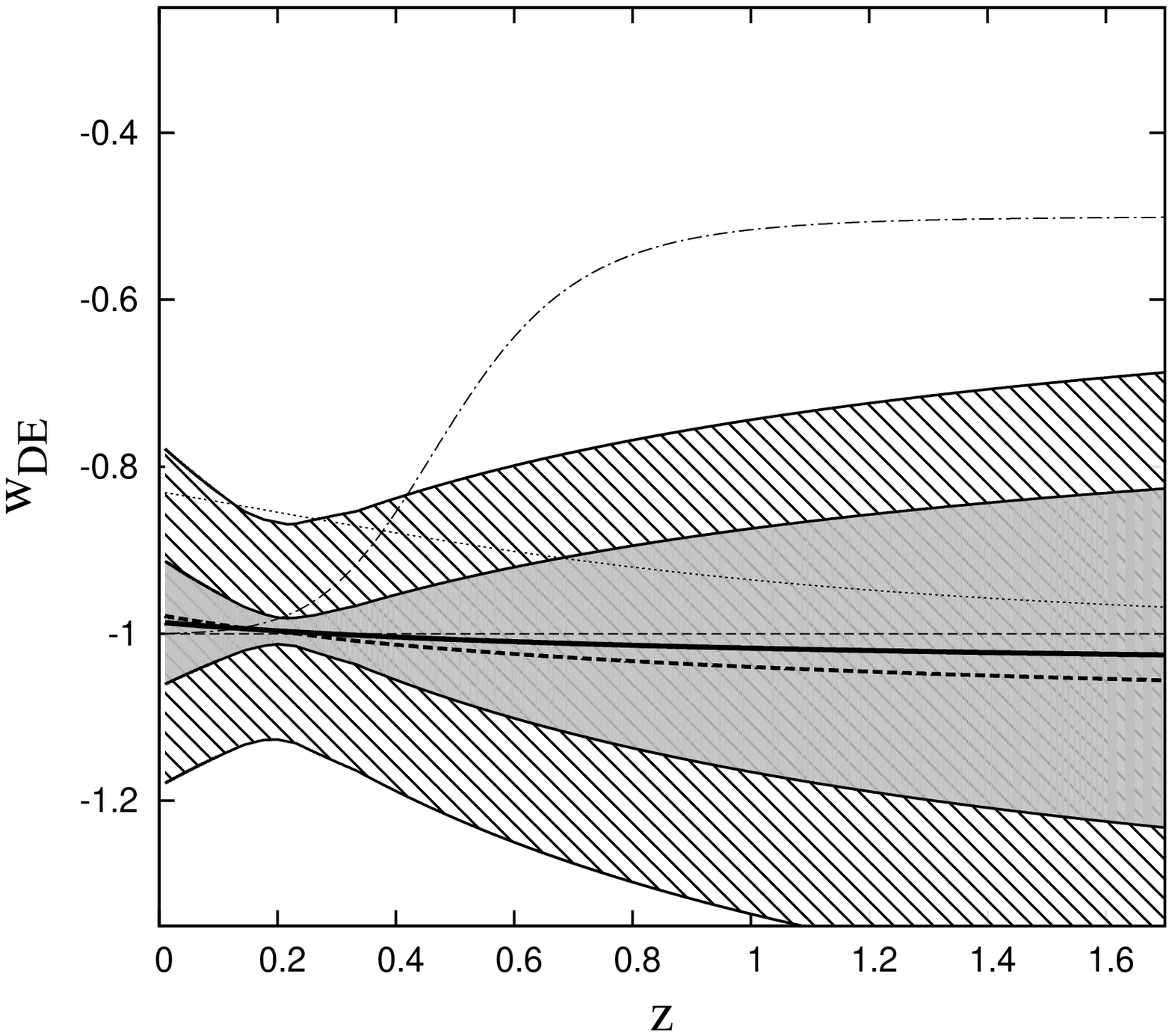} \\ [0.4cm]
\mbox{\bf (c)} & \mbox{\bf (d)}
\end{array}$
\end{center}
\caption{
Variation of different cosmological parameters with redshift for
dataset II A. Panel (a) shows the quantity $\omm(z)$, panel (b) shows
$\omde(z)$, panel (c) shows $q(z)$ and panel (d) shows $\w(z)$. The
shaded contours represent the $2\sigma$ confidence levels, the thick
solid lines represent the best-fit, the thick dashed lines in panels
(b) and (d) represent the best-fit when $\omt$ is marginalized over
. The dashed, dotted and dot-dashed lines represent the true model for
datasets II A, B and C respectively. In panels (b) and (d), the grey
solid contours represent the $2\sigma$ confidence levels when $\omt$
is known exactly, while the hatched contours are marginalized over
$\omt =0.27 \pm 0.03$.  }
\label{fig:dat2}
\end{figure*}

To study the efficacy of the different parameters in quantifying \de,
we use both geometrical and physical parameters-- we select the
geometrical parameters $\omm(z)$ and $q(z)$, and the physical
parameters $\omde(z)$ and $\w(z)$. For each parameter we need to
choose a fitting function which will be used in the likelihood
analysis. We experimented with different fitting functions, \eg
comparing the polynomial expansion in redshift $z$ with that in scale
factor $a$ for $\w(z)$; and in each case we choose the fitting
function which results in the best reconstruction of the parameter of
interest. The fitting functions chosen for the different parameters
are--
\ber
\omm(z) &=& \omm_0 + \omm_1 (1+z) \\
q(z) &=& q_0 + \frac{q_1 z}{1+z} \\
\omde(z) &=& (1-\omt-\Omega_1-\Omega_2) \nonumber\\
&&+ \Omega_1 (1+z) + \Omega_2 (1+z)^2 \nonumber\\
&=& \Omega_0 + \Omega_1 z + \Omega_2 z^2 \\
\w(z)&=& w_0 + \frac{w_1 z}{1+z} \,\,,
\eer
of which the fitting function for $q(z)$ was first introduced in
\cite{qparm}, that for $\omde(z)$ was introduced in \cite{state}, and
that for $\w(z)$ in \cite{CPL}. We use these fitting function for the
likelihood estimation to obtain confidence levels on the cosmological
parameters, which can then be used to discriminate between different
theoretical models of dark energy.

For the geometrical parameters of \de, no further information is
necessary for performing the analysis. However, for the physical
parameters $\omde(z)$ and $\w(z)$, information is required on the
matter density. We therefore do the analysis for the physical
parameters by first fixing the value of matter density to $\omt =
0.27$ (which is the true value of $\omt$ for the datasets II A, B, C,
and is the expected value today from large scale structure), then by
marginalizing over the matter density using $\omt = 0.27 \pm 0.03$
\cite{sdss}. The results for the second case are expected to be worse
for the physical parameters, while for the geometrical parameters
there is no difference since they do not depend on $\omt$.

\section{Results}\label{res}

\begin{figure*}
\begin{center}
$\begin{array}{c@{\hspace{1in}}c}
\epsfxsize=2.4in
\epsffile{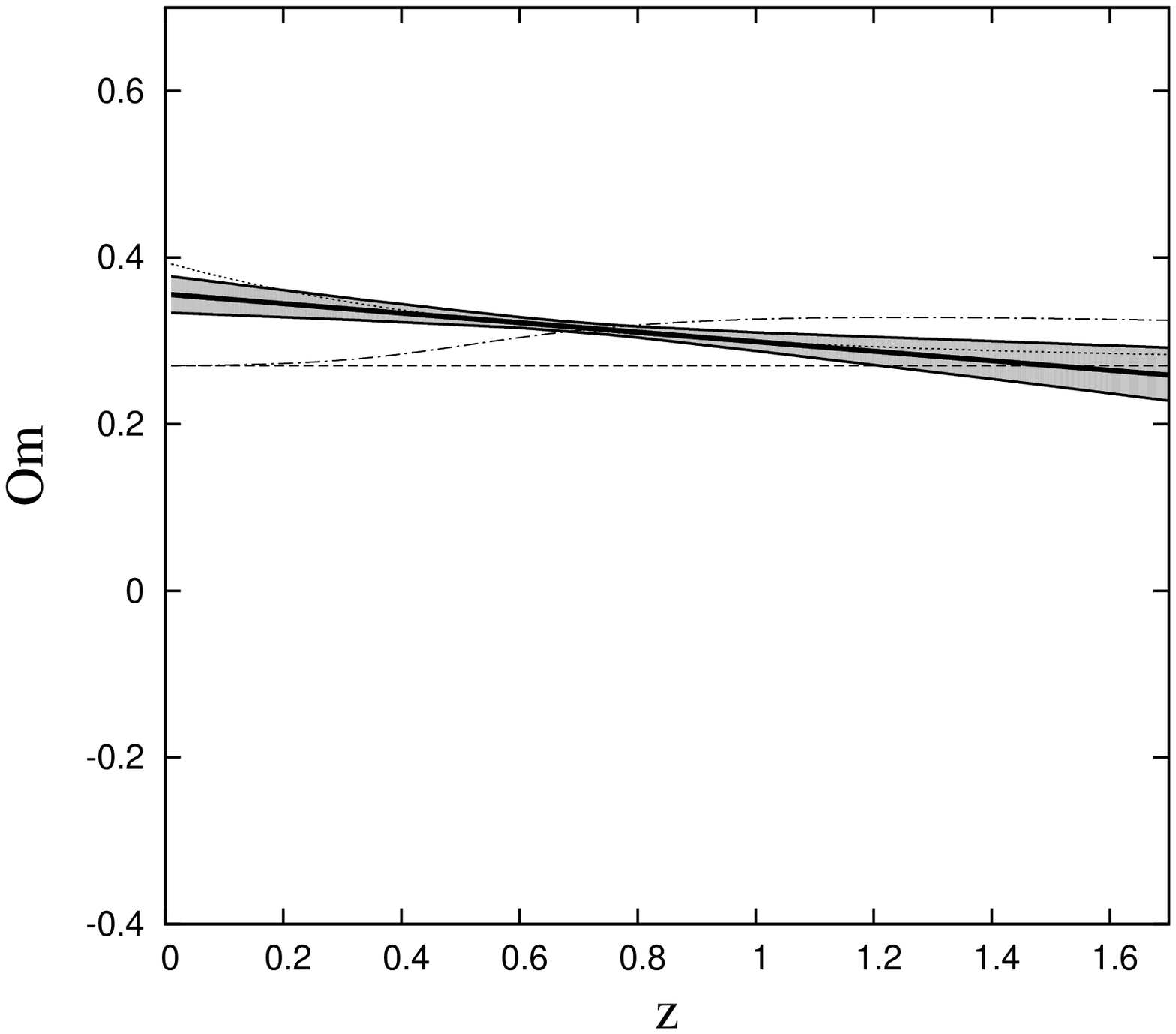} &
	\epsfxsize=2.4in
	\epsffile{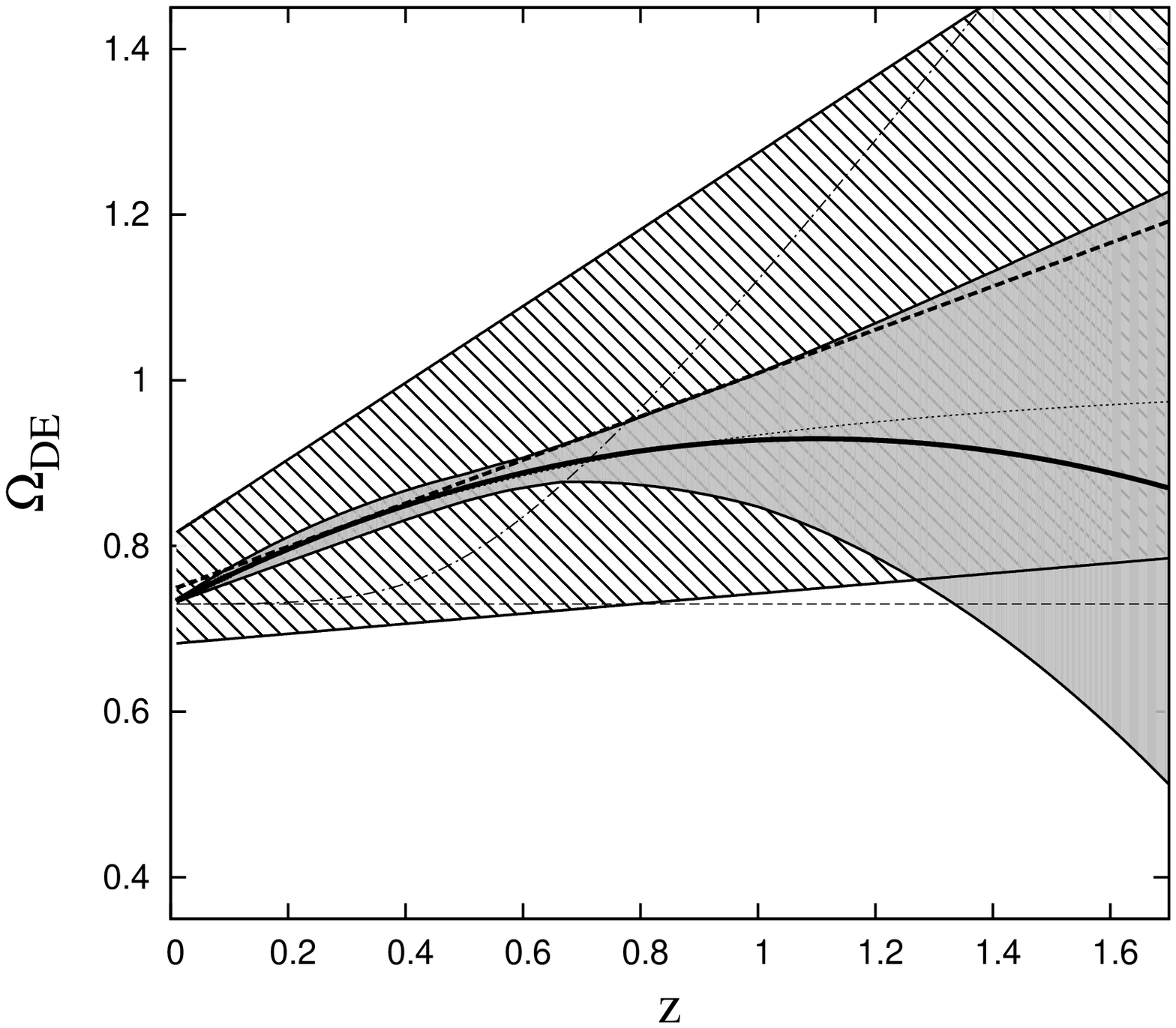} \\ [0.4cm]
\mbox{\bf (a)} & \mbox{\bf (b)}
\end{array}$
$\begin{array}{c@{\hspace{1in}}c}
\epsfxsize=2.4in
\epsffile{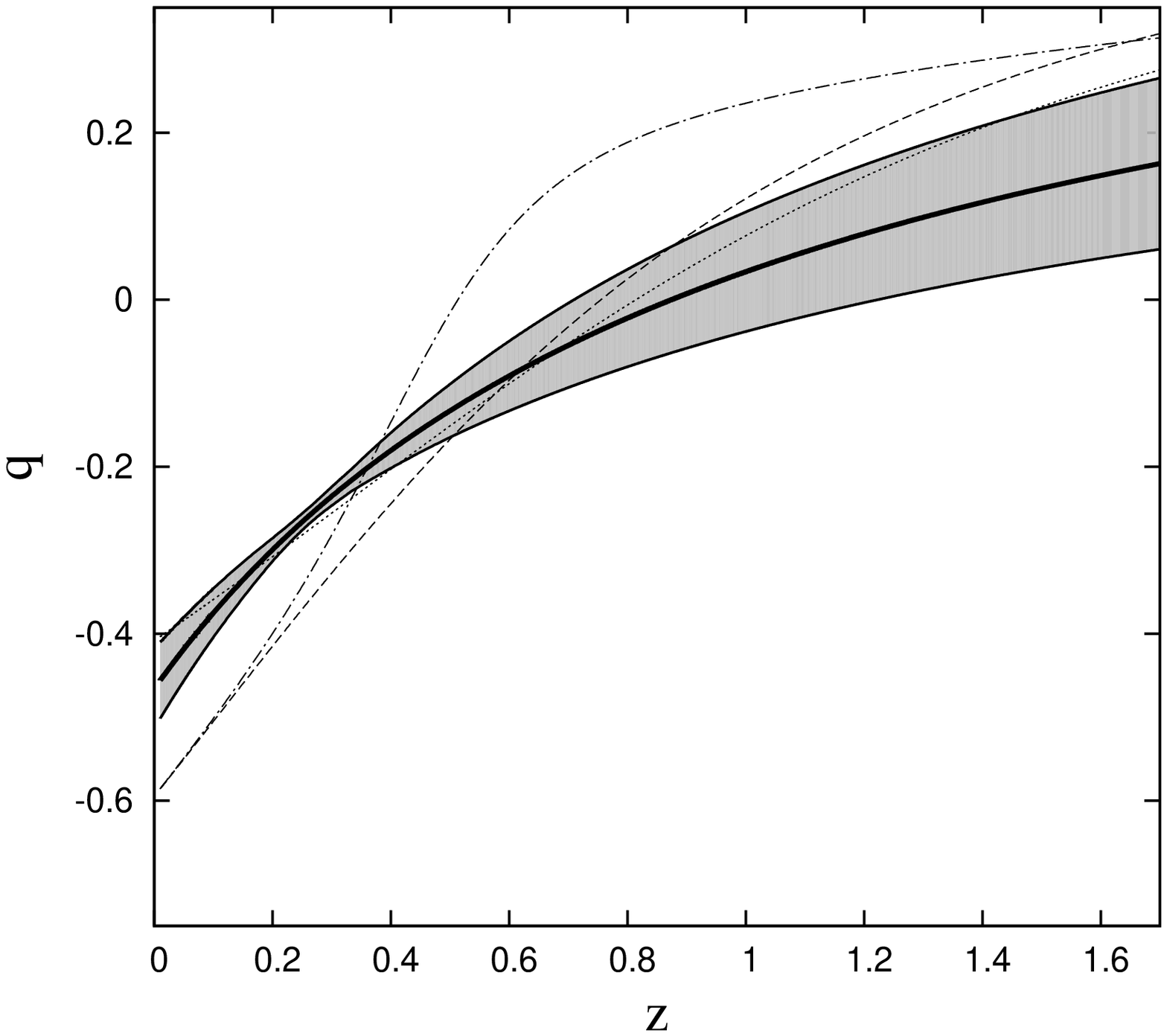} &
	\epsfxsize=2.4in
	\epsffile{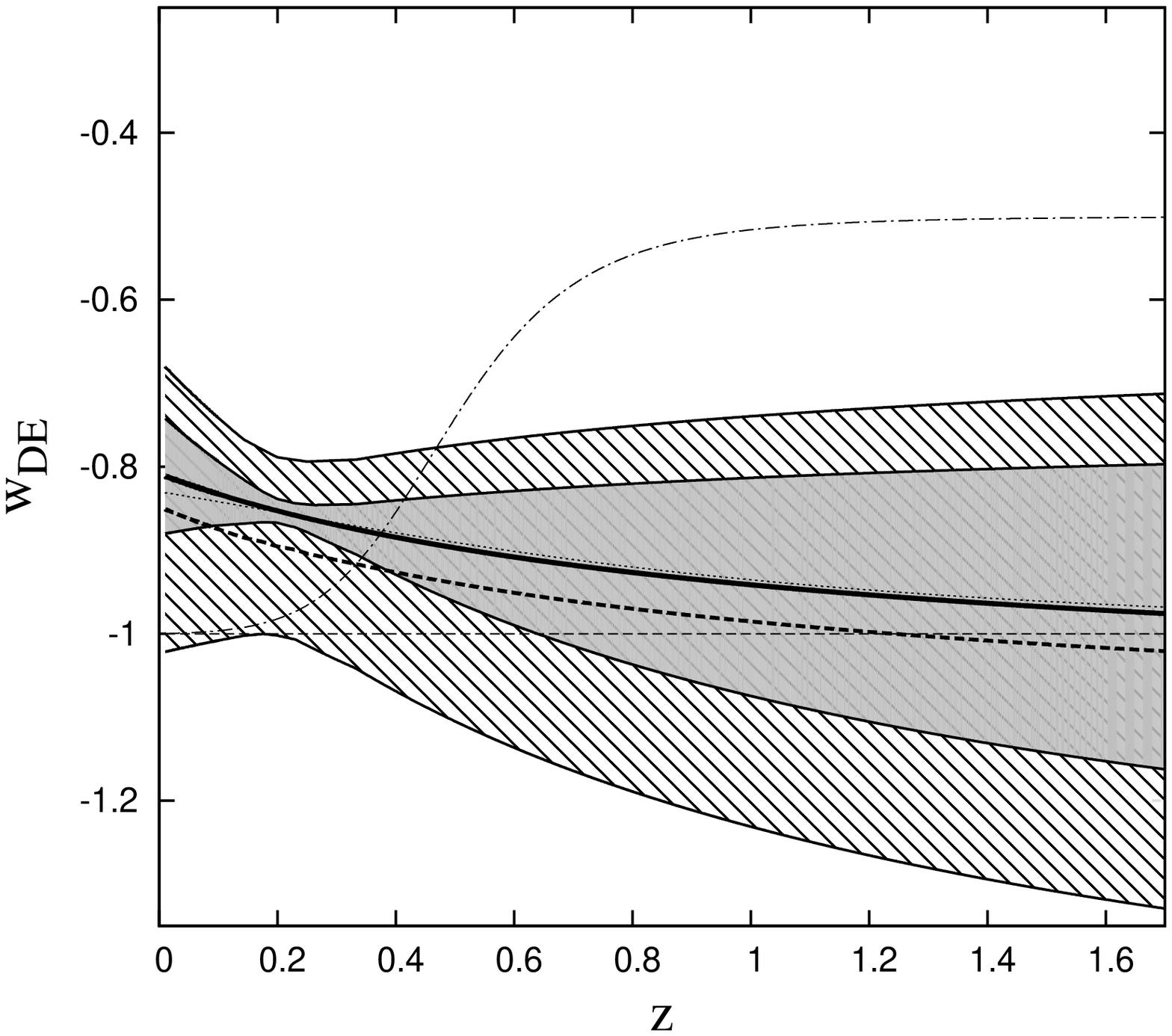} \\ [0.4cm]
\mbox{\bf (c)} & \mbox{\bf (d)}
\end{array}$
\end{center}
\caption{
Variation of different cosmological parameters with redshift for
dataset II B. Panel (a) shows the quantity $\omm(z)$, panel (b) shows
$\omde(z)$, panel (c) shows $q(z)$ and panel (d) shows $\w(z)$. The
shaded contours represent the $2\sigma$ confidence levels, the thick
solid lines represent the best-fit, the thick dashed lines in panels
(b) and (d) represent the best-fit when $\omt$ is marginalized over
. The dashed, dotted and dot-dashed lines represent the true model for
datasets II A, B and C respectively. In panels (b) and (d), the grey
solid contours represent the $2\sigma$ confidence levels when $\omt$
is known exactly, while the hatched contours are marginalized over
$\omt =0.27 \pm 0.03$. }
\label{fig:dat3}
\end{figure*}

\begin{figure*}
\begin{center}
$\begin{array}{c@{\hspace{1in}}c}
\epsfxsize=2.4in
\epsffile{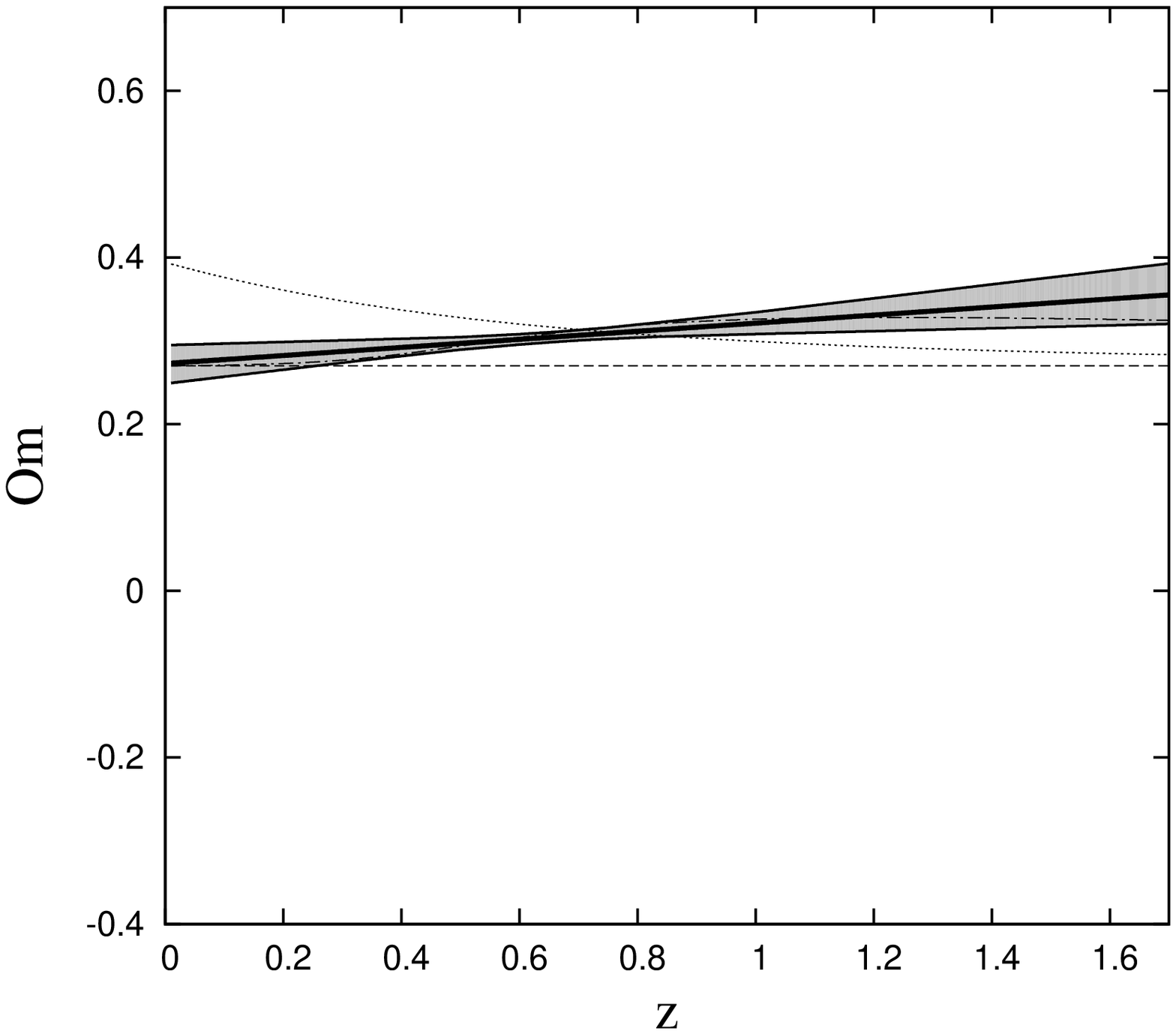} &
	\epsfxsize=2.4in
	\epsffile{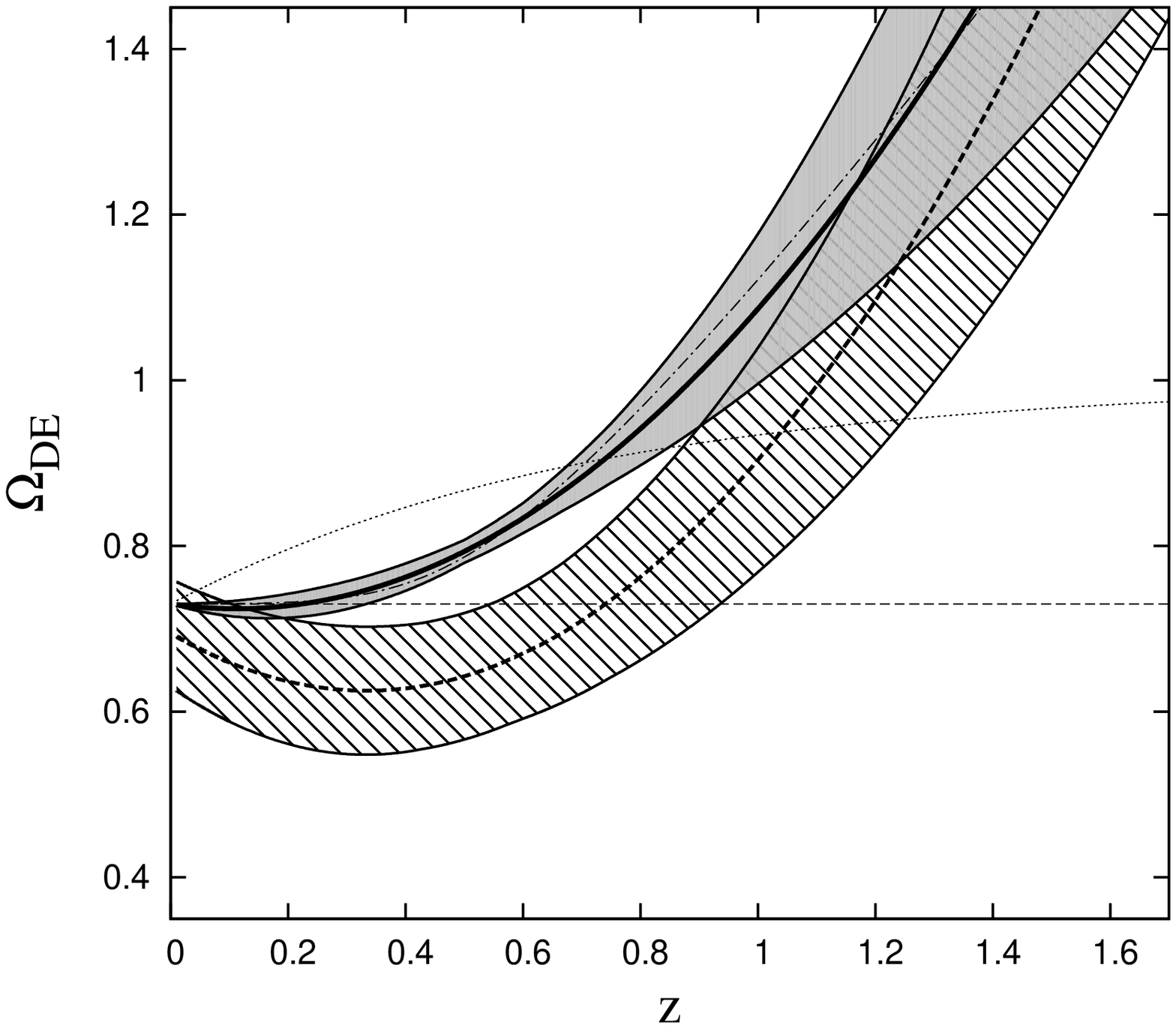} \\ [0.4cm]
\mbox{\bf (a)} & \mbox{\bf (b)}
\end{array}$
$\begin{array}{c@{\hspace{1in}}c}
\epsfxsize=2.4in
\epsffile{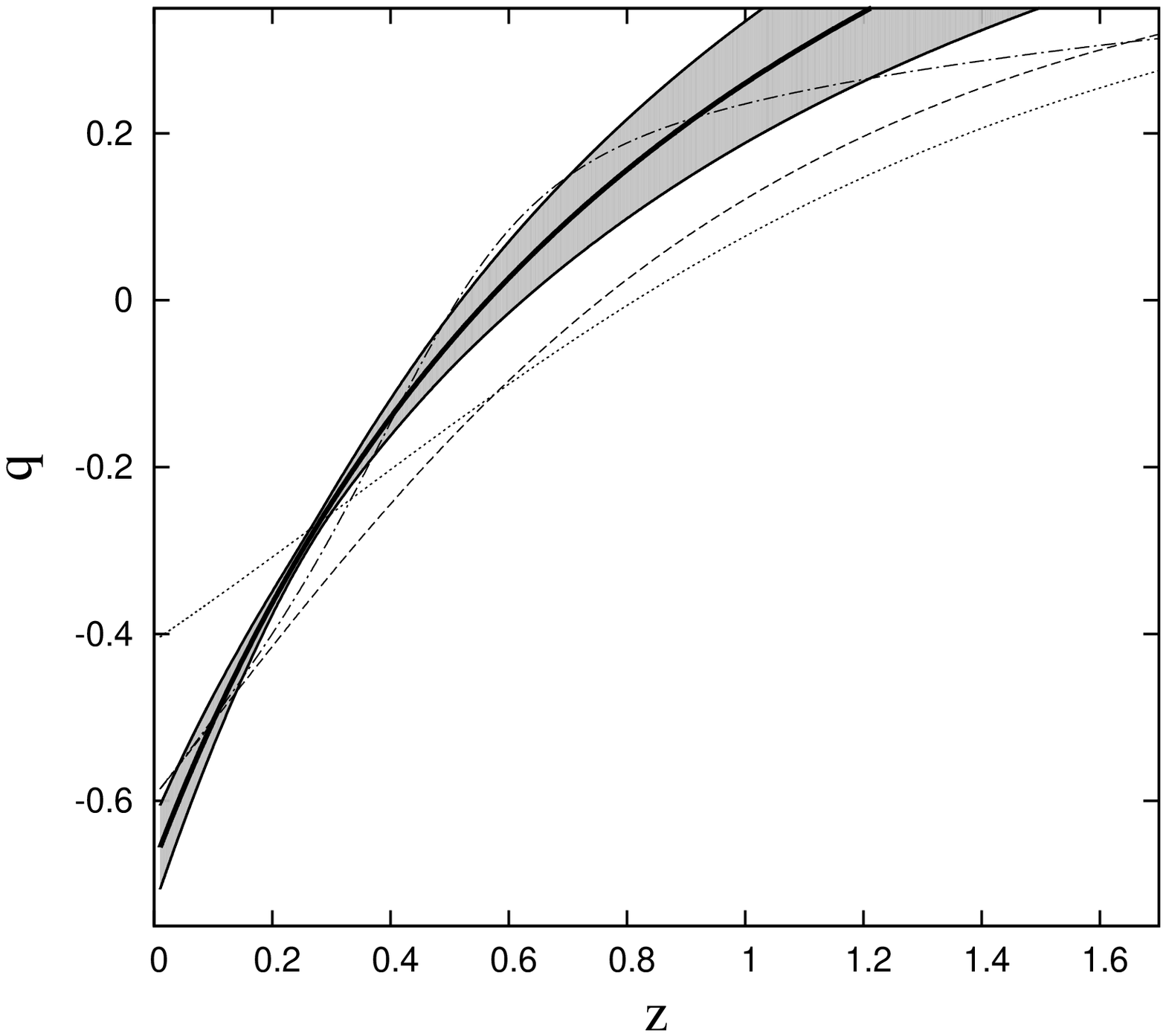} &
	\epsfxsize=2.4in
	\epsffile{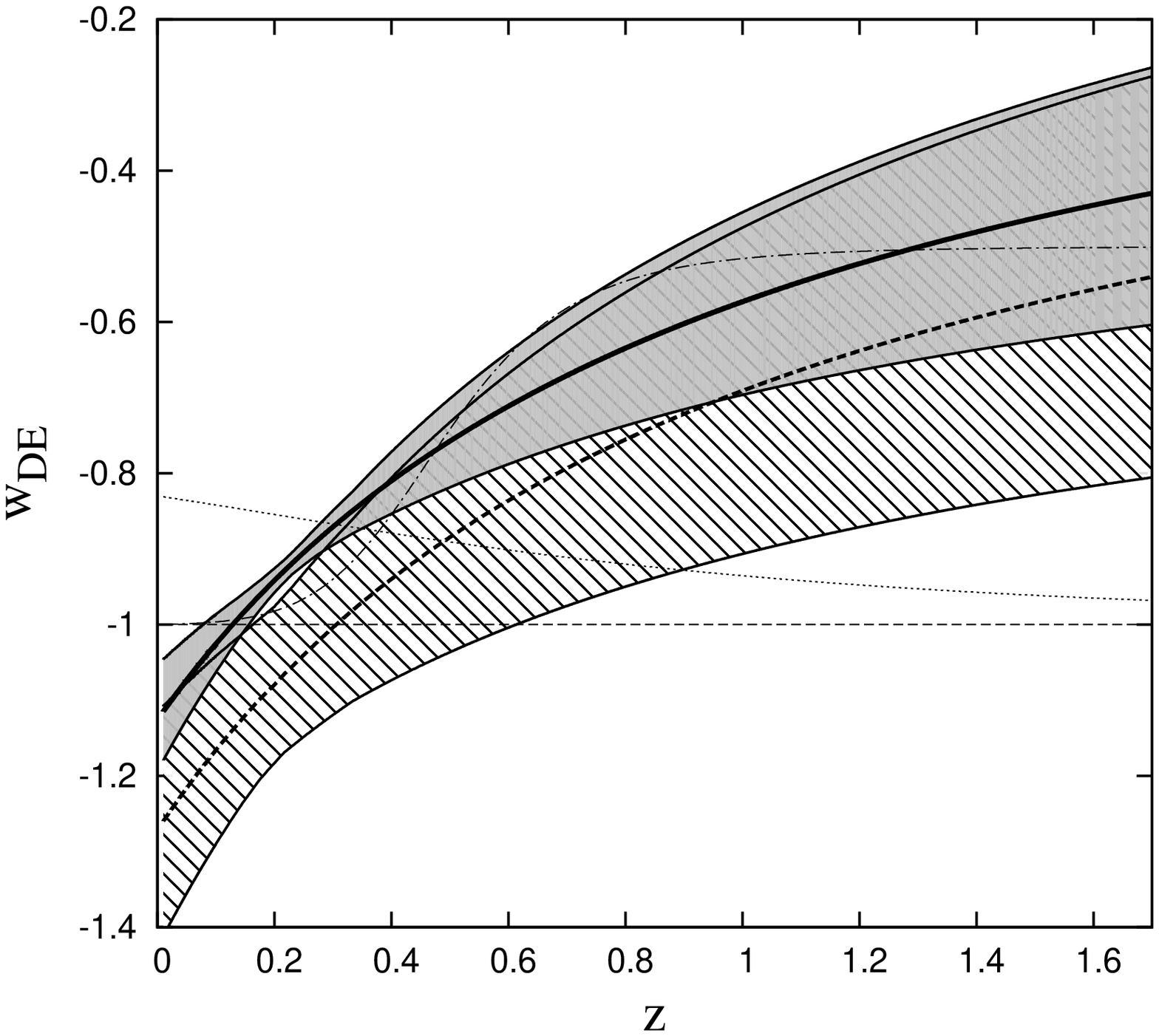} \\ [0.4cm]
\mbox{\bf (c)} & \mbox{\bf (d)}
\end{array}$
\end{center}
\caption{
Variation of different cosmological parameters with redshift for
dataset II C. Panel (a) shows the quantity $\omm(z)$, panel (b) shows
$\omde(z)$, panel (c) shows $q(z)$ and panel (d) shows $\w(z)$. The
shaded contours represent the $2\sigma$ confidence levels, the thick
solid lines represent the best-fit, the thick dashed lines in panels
(b) and (d) represent the best-fit when $\omt$ is marginalized over
. The dashed, dotted and dot-dashed lines represent the true model for
datasets II A, B and C respectively. In panels (b) and (d), the grey
solid contours represent the $2\sigma$ confidence levels when $\omt$
is known exactly, while the hatched contours are marginalized over
$\omt =0.27 \pm 0.03$. }
\label{fig:dat4}
\end{figure*}

\begin{table*}
\begin{center}
\caption{
\scriptsize $\chi^2_{\rm min}$ for the different parameter
reconstructions using the three simulated datasets, the efficiency
criterion (\ref{eq:eff}) which defines how well the model reconstructs
the truth, and the discrimination criterion (\ref{eq:disc}), which
defines how well the model discriminates from the other two truths
considered. }
\begin{tabular}{c|ccccc}
\hline
Dataset & $\omt$ & Parameterization & $\chi^2_{\rm min}$ & $\delta^2_p=\sum (p_{\rm true}-p_{\rm fit})^2 \sigma^2_{p({\rm fit})}$ & $\ \frac{\delta_p^2({\rm True \ Model})}{\delta_p^2({\rm Nearest \ Model})}$ \\
\hline
& & & & &\\
 II A & $0.27$ & $\omm(z) = \omm_0 + \omm_1 (1+z)$ & $2349.48$ & $10^{-5}$ & $0.001$\\
& & $q(z) = q_0 - q_1 z/(1+z)$ & $2349.77$ & $0.011$ & $0.668$\\
& & $\omde(z) =  \Omega_0 + \Omega_1 z + \Omega_2 z^2$ & $2349.45$ & $0.019$ & $0.009$\\
& & $w(z) = w_0 - w_1 z/(1+z)$ & $2349.44$ & $0.014$ & $0.010$\\
& & & & &\\
& $0.27 \pm 0.03$ &  $\omde(z) =  \Omega_0 + \Omega_1 z + \Omega_2 z^2$ & $2349.03$ & $0.164$ & $0.882$\\
& & $w(z) = w_0 - w_1 z/(1+z)$ & $2349.26$ & $0.173$ & $1.395$\\
\hline
& & & & &\\
 II B & $0.27$ & $\omm(z) = \omm_0 + \omm_1 (1+z)$ & $2349.75$ & $0.0002$ & $0.007$\\
& & $q(z) = q_0 - q_1 z/(1+z)$ & $2350.08$ & $0.047$ & $0.652$\\
& & $\omde(z) =  \Omega_0 + \Omega_1 z + \Omega_2 z^2$ & $2349.54$ & $0.119$ & $0.010$\\
& & $w(z) = w_0 - w_1 z/(1+z)$ & $2349.45$ & $0.042$ & $0.009$\\
& & & & &\\
& $0.27 \pm 0.03$ &  $\omde(z) =  \Omega_0 + \Omega_1 z + \Omega_2 z^2$ & $2349.18$ & $0.292$ & $1.002$\\
& & $w(z) = w_0 - w_1 z/(1+z)$ & $2349.37$ & $0.201$ & $1.292$\\
\hline
& & & & &\\
II C & $0.27$ & $\omm(z) = \omm_0 + \omm_1 (1+z)$ & $2349.75$ & $0.0001$ & $0.006$\\
& & $q(z) = q_0 - q_1 z/(1+z)$ & $2349.50$ & $0.104$ & $0.479$\\
& & $\omde(z) =  \Omega_0 + \Omega_1 z + \Omega_2 z^2$ & $2349.46$ & $0.039$ & $0.007$\\
& & $w(z) = w_0 - w_1 z/(1+z)$ & $2349.97$ & $0.077$ & $0.023$\\
& & & & & \\
& $0.27 \pm 0.03$ &  $\omde(z) =  \Omega_0 + \Omega_1 z + \Omega_2 z^2$ & $2349.19$ & $1.981$ & $0.452$\\
& & $w(z) = w_0 - w_1 z/(1+z)$ & $2348.96$ & $1.163$ & $0.356$\\
\hline
\end{tabular}\label{tab:chi2}
\end{center}
\end{table*}

We first study the currently available data. The results for the Union2
dataset, \ie, dataset I, using $\omt = 0.27$, are shown in
figure~\ref{fig:dat1}. We see that the $2\sigma$ confidence levels for
all four parameters are extremely large, and the three very different
\de~models proposed in datasets II A, B, C all fall within these
confidence levels for at least a large part of the redshift
range. Therefore, with the current data, at $95 \%$ confidence, three
widely varying models of \de~cannot be distinguished very well from
each other using different cosmological parameters, even if the matter
density were known exactly.

Since the current data cannot be used to effectively constrain
theoretical models of \de, we now look at the constraints that may be
obtained from future observations. Dataset II A is a JDEM-like
simulated dataset for which the true underlying model is the
\cc. Figure~\ref{fig:dat2} shows the $2\sigma$ confidence levels on the
different parameters for this dataset. We see that the parameter
$\omm$ has the lowest errors and represents the true model quite
accurately. The deceleration parameter $q$ also has narrow errors,
however, since the three cosmological models considered are much
closer in $q$-space than in the other parameter spaces, even with
these narrow errors, $q$ cannot distinguish between the \de~models of
dataset II A and dataset II B. Both the equation of state $\w$ and the
energy density $\omde$ have somewhat larger errors but do a better job
as compared to the deceleration parameter of constraining the true
model as well as distinguishing it from the other two \de~models
considered, when exact value of $\omt$ is known.  The parameter
$\omde$ has especially narrow error bars at low redshifts, but after a
redshift of $z \sim 0.7$ the error bars become larger. When value of
$\omt$ is not known exactly this leads to additional uncertainty in
the results, as well as the possibility of bias. A higher value of
$\omt$ achieves a luminosity distance similar to that obtained by
\de~with an increasing $w$, therefore there is a degeneracy between
$\omt$ and the \de~parameters, leading to the larger error bars on the
parameters $\w$ and $\omde$.

Next, we look at dataset II B, which has an underlying model of
quintessence with a slowly varying equation of
state. Figure~\ref{fig:dat3} shows the $2\sigma$ confidence levels for
the different parameters for this dataset. The results are very
similar to that for dataset II A. The parameter $\omm$ reconstructs
quite accurately with narrow error bars at high redshifts, but is
slightly less effective at $z = 0$. The parameter $q$ reconstructs
accurately at low $z$ at less so at high $z$. The parameter $\omde$ is
excellent at reconstruction at low redshift when $\omt$ is known,
while $w$ reconstructs reasonably as well when $\omt$ is known.  When
$\omt$ is not known accurately, the reconstruction has a bias for
$\omde$ and $w$, and the error bars are larger.

Dataset II C uses a model with a stronger variation of the equation of
state, from $\w = -1$ today to $\w = -0.5$ at $z = 1$. The results
for this dataset are shown in figure~\ref{fig:dat4}. We see that both the
deceleration parameter $q$ and the equation of state $\w$ are unable
to reconstruct the true model (dot-dashed lines) accurately, because
the parameterization chosen is not general enough to reconstruct the
true model accurately. Even when $\omt$ is known, the equation of
state appears to favour $\w < -1$ at $z = 0$ at $2\sigma$, which would
lead to an extremely different model of dark energy violating the Weak
Energy Condition. The \de~density $\omde$ does better in
reconstructing the true model when $\omt$ is known, especially at low
redshift. This is because the \de~density, which is obtained from the
first derivative of the SNe data, has usually less sharp variations
than the equation of state or deceleration parameter, which are
related to the second derivative of the data. However, when $\omt$ is
not known exactly, $\omde$ performs much worse, especially at low
redshift, because of the degeneracy between $\omde$ and the matter
density. As in the previous cases, $\omm$ appears to capture the
behaviour of this model very well.

As we see from the
figures~\ref{fig:dat2},~\ref{fig:dat3},~\ref{fig:dat4}, the different
parameters perform very differently in reconstructing the models
chosen.  We quantify the efficiency of the different parameters using
a distance criterion which calculates the distance of the true model
from the best-fit, weighted by the $2\sigma$ errors obtained by the
fitting procedure. The lower this value, the more accurate the
reconstruction, and the narrower the confidence levels on the
parameter. A large value of this parameter could mean either very
large errors, which lowers the discriminatory power of the parameter,
or a biased fit that is far away from the true result. We also
quantify the distance criterion for the other two models in each case
and take the ratio of this quantity for the true model to that for the
nearest of the other models, since this highlights the discriminatory
power of the parameter. The lower this value, the better the parameter
at discriminating other models from the true model. We note that this
quantity is of course dependent on the models chosen, however, for the
three models studied here, it can give a comparative assessment of the
discriminatory power of the parameters considered. For a different set
of models, the value of this quantity would obviously change, but the
comparison between the cosmological parameters would be similar,
especially since the three models chosen are quite different from each
other in behaviour. For a parameter $p$, the efficiency criterion is
defined as-- 
\beq\label{eq:eff} 
\delta^2_p=\sum_i (p_{i, {\rm true}}-p_{i, {\rm fit}})^2
\sigma^2_{p(i, {\rm fit})}\,\,, 
\eeq
and the discrimination ratio is defined as-- 
\beq\label{eq:disc}
\frac{\delta_p^2({\rm True \ Model})}{\delta_p^2({\rm Nearest \ Model})} \,\,, 
\eeq 
where $\sigma^2_{p(i, {\rm fit})}$ is the $2\sigma$ error on the
reconstructed parameter.

The above quantities are calculated in Table~\ref{tab:chi2} for the
different parameters for all three models. We see that for the
different parameters, the $\chi^2$ on the data is quite similar for
all cases, showing that they all reconstruct successfully in the data
space. However, since the parameter space is linked to the data space
by one or more differentiations, the reconstruction of the actual
parameters, quantified by $\delta^2_p$ (\ref{eq:eff}), is quite
different for the different parameters. The quantity $\omm$ performs
consistently well in reconstructing the \de~models considered, and has
the lowest values of $\delta^2_p$. The deceleration parameter $q$
performs well for Models A, B, but does not do so well for Model C
which is a model not well characterized by the parameterization chosen
for it. The \de~density and the \de~equation of state both work
reasonably when the matter density is known, however, their
performance degenerates when uncertainty is introduced in $\omt$,
especially in Model C which is the most strongly varying. The
discrimination criterion (last column of table) shows how well the
parameter can discriminate the true model from the other two models
considered. We see that the parameter $\omm$ performs very well on
this criterion as well. The deceleration parameter, which performed
reasonably well on $\delta^2_p$, does not perform as well on this
criterion, because the different cosmological models are quite close
together when represented by this quantity. Thus the deceleration
parameter is not very useful when it comes to choosing one \de~model
over the others. The \de~density $\omde$ and the equation of state
$\w$ both perform reasonably when $\omt$ is known exactly, but once
again their performance degenerates when priors are put on $\omt$,
since the error bars increase. Overall, in both the criteria
considered, $\omm$ appears to perform the best, while $\omde$ performs
well at low $z$ provided $\omt$ is known, and $w$ performs moderately
well when $\omt$ is known.

\section{Conclusions}\label{concl}

We reconstruct four different cosmological parameters to examine their
potential in extracting information from observations about \de. We
find that the parameter $\omm$, which is constructed from the Hubble
parameter, has the narrowest errors, and gives the most accurate
reconstruction. We also see that results for the physical parameters
$\omde, \w$ are extremely dependent on knowledge of the matter
density, and without very tight bounds on matter density, physical
parameters of \de~may not perform well in reconstruction \de. The
deceleration parameter reconstructs quite well, but has low
discriminatory powers, since different models of \de~have very similar
values for this parameter.

Overall it appears that the parameters constructed out of the first
derivative of the data, \ie the Hubble parameter, are somewhat better
poised to give information about the nature of dark energy than the
second derivatives, and geometric parameters of \de, such as $\omm$,
are better at reconstructing \de~since they are not biased by lack of
knowledge about the matter density, provided they are constructed in
such a way that they can discriminate between different models of
\de. To obtain information about \de~from the physical parameters of
\de, it is important to have independent sources of observation of the
matter density. Thus, for a parameter to obtain maximum information
out of the data, it works better if it is constructed out of the first
derivative of the data or less, does not depend on other physical
parameters such as the matter density, and is constructed such that
different \de~models can be easily discriminated from each other with
the parameter.

We note here that the reconstruction methods considered here are that
of parametric reconstruction, non-parametric reconstruction of
cosmological parameters would suffer less from biases in
reconstruction such as those Model C. However, typically these methods
have higher errors, and also the issues of degeneracy with other,
non-\de~parameters exist for these methods as well. Therefore if a
simple cosmological parameter, like $\omm$, can be constructed which
performs well for a large class of models, both in reconstructing and
in discriminating between models, and is also independent of other
non-\de~parameters such as $\omt$, it would be extremely useful for
understanding the nature of \de.

\acknowledgments

UA acknowledges support from the LDRD program at Los Alamos National
Laboratory.  AVP would like to thank the Material Designs Institutes
Summer School, 2009 at LANL and the hospitality of the ISR-1 group of
LANL. We acknowledge useful discussions with K. Heitmann and S. Habib.

\end{document}